\documentclass{cimento}

\usepackage{graphicx}  
\usepackage{amsmath}

\title{Quarkonium spectral functions in a bulk-viscous quark-gluon plasma}
\author{Lata Thakur\from{ins:x}
	\thanks{thakurphyom@gmail.com, present address: Department of Physics and Institute of Physics and Applied Physics, Yonsei University, Seoul 03722, Korea}
	\atque
	Yuji~Hirono\from{ins:x}\thanks{yuji.hirono@gmail.com, present address: Department of Physics, Kyoto University, Kyoto 606-8502, Japan}}
\instlist{\inst{ins:x} Asia Pacific Center for Theoretical Physics, 
	Pohang, Gyeongbuk 37673, Republic of Korea}
\begin{document}
	
	\maketitle
	
\begin{abstract}
	We study the interplay of non-equilibrium properties of a quark-gluon plasma (QGP) and heavy quarkonia.
	For this purpose, we compute the quarkonium spectral functions in a bulk-viscous QGP. 
	We take into account the bulk viscous nature of the medium by modifying the distribution functions of thermal quarks and gluons. 
	This modification affects the dielectric permittivity, 
	which is used to calculate the in-medium heavy quark potential.
	With this modified complex potential, we calculate the quarkonium spectral functions and extract their physical properties. 
	We discuss the impact of bulk viscosity on quarkonia properties such as 
	decay widths and binding energies. 
	We also estimate the relative production yield of  $ \psi' $ to $ J/\psi $  and discuss the bulk viscous effects on it.
\end{abstract}

\section{Introduction}

Quarkonia play a pivotal role as invaluable probes in the exploration of quark-gluon plasma (QGP) formation in heavy-ion collision experiments. 
The observation of quarkonium state suppression serves as an indicator of QGP formation, a phenomenon initially proposed by Matsui and Satz in their seminal work~\cite{Matsui:1986dk}.
Notably, recent researches have pointed out that the bulk viscosity of the QCD medium is going to be enhanced near the QCD critical point, which has garnered heightened attention, especially within the context of the beam energy scan program~\cite{Kharzeev:2007wb,Karsch:2007jc,Ryu:2015vwa,Bzdak:2019pkr}.
This contribution centers on the exploration of the relationship between the bulk viscosity of the medium and quarkonium properties via the evaluation of quarkonium spectral functions in a bulk viscous medium~\cite{Thakur:2021vbo}.
We estimate the impact of bulk viscosity on the physical properties of quarkonia, such as decay widths and binding energies. 
We also evaluate the relative production yield of $ \psi' $ to $ J/\psi $ ratio, which is experimentally measurable.

\section{ Computation of quarkonia spectral functions in the presence of bulk viscosity }

Let us outline the computational procedure of spectral functions in the presence of bulk viscosity. 
We first introduce the bulk viscous correction
as a deformation of the distribution function
of thermal particles as~\cite{Du:2016wdx}
\begin{eqnarray}
	f(k) 
	=
	f_{0}(\tilde{k}) +\delta_{\rm bulk}f(\tilde{k}) 
	=
	f_{0}(\tilde{k})\left(1+\frac{m^{2}\Phi}{2T\sqrt{k^2+m^2}}(1\pm 	f_{0}(\tilde{k})) \right), 
	\label{fk}
\end{eqnarray}
where $\tilde{k}\equiv \frac{1}{T}\sqrt{k^{2}+m^{2}}  $, $ f_{0}(\tilde{k}) $ is the equilibrium distribution, 
$\Phi$ is a parameter quantifying 
the strength of bulk-viscous correction, 
and $ m $ is the quasiparticle mass, which depends upon temperature and chemical potential~\cite{Peshier:1995ty}. 
Based on the modified distribution function~(\ref{fk}), the total retarded self energy is computed as 
\begin{equation}
	\Pi_{R}(P)
	=-\bar{m}^2_{D,R}\left(1-\frac{p^{0}}{2p}\ln\frac{p^{0}+p+i\epsilon}{p^{0}-p+i\epsilon}\right),
	\label{PiR}
\end{equation}
where $\bar{m}^2_{D,R}=m^2_{D,R}+\delta m^2_{D,R}$ is the retarded Debye mass in the presence of bulk viscous correction. 
The non-equilibrium correction $\delta m^2_{D,R}$ is given by 
\begin{equation}
	\delta m^2_{D,R}=\frac{g^2 T^2}{2}\frac{\bar{m}^2}{\pi^2 }\Phi\left[ 2N_c \left( \frac{1}{e^{\bar{m}}- 1}  \right) + N_f \left( \frac{1}{1+e^{\bar{m}- \bar{\mu}}}+\frac{1}{1+e^{\bar{m}+\bar{\mu}} }\right)  \right] ,
\end{equation}
where $ \bar{m}=m/T $ and $ \bar{\mu}=\mu/T $. 
Similarly, the Debye mass of the symmetric propagator can 
be computed as 
$\bar{m}^2_{D,S}=m^2_{D,S}+\delta m^2_{D,S}$, 
where the first term is the equilibrium contribution. 
Using these quantities, we obtain the dielectric permittivity 
$\epsilon(p) $ as 
\begin{equation}
	\epsilon^{-1}(p) = 
	\frac{p^2}{p^2 + \bar{m}^2_{D,R}}
	- i  
	\frac{\pi T p  \, 
		\bar{m}^2_{D,S}  
	}{ (p^2 + \bar{m}^2_{D,R})^2} . 
	\label{eq:epsilonphi}
\end{equation}
\begin{figure}
	\includegraphics[width=13cm]{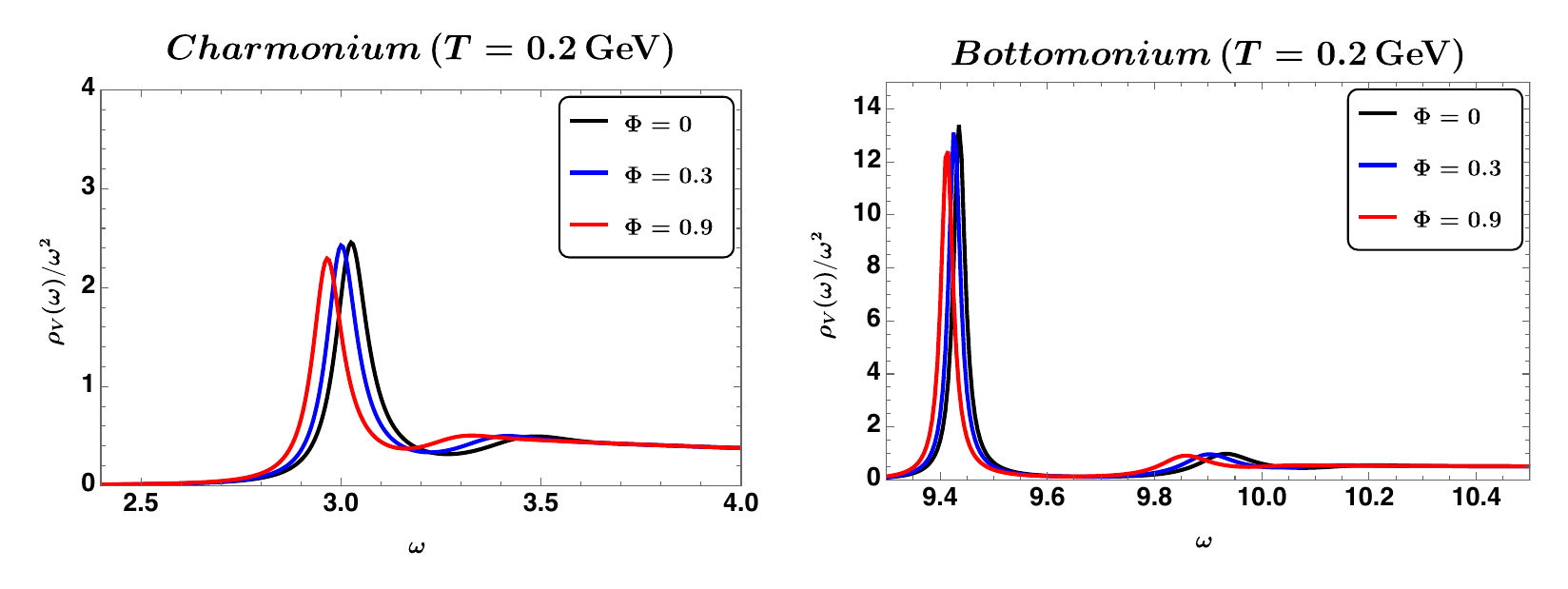}     
	\caption{S-wave	vector channel spectral functions of charmonium(left) and bottomonium(right) for different values of $ \Phi $ at $T=0.2 $ GeV}.
	\vspace{-1.1cm}
	\label{SFvsw}
\end{figure}
%

We use the dielectric permittivity to compute the in-medium heavy quark complex potential by modifying the Fourier transform of the Cornell potential~\cite{Thakur:2021vbo,Thakur:2020ifi,Thakur:2013nia}.
The real and imaginary parts of the potential are
\begin{align}
	{\Re}V(r,T,\Phi)
	&=	-\alpha \, 
	\bar{m}_{D,R}
	\left(\frac{e^{-\bar{m}_{D,R}\, r}}{\bar{m}_{D,R}\, r}+1\right) 
	+ 
	\frac{2\sigma}{
		\bar{m}_{D,R} 
	}\left(\frac{e^{-\bar{m}_{D,R}\,{r}}-1}{\bar{m}_{D,R}\,r}+1\right),
	\label{eq:pot-realphi}
	\\
	{\Im}V(r,T,\Phi)  
	&=
	-\alpha \lambda T 
	\, \phi_2  (\bar{m}_{D,R}\, r)
	- 
	\frac{ 
		2\sigma \lambda T 
	} {
		\bar{m}^2_{D, R} 
	}
	\, 
	\chi( \bar{m}_{D, R}\, r ),
	\label{eq:pot-imaginary}
\end{align}
where  $\lambda	\equiv  \bar m^2_{D, S} / \bar m^2_{D, R} $, 
and $\alpha$ is the strong coupling constant. 
%
\begin{figure}
	\includegraphics[width=14cm]{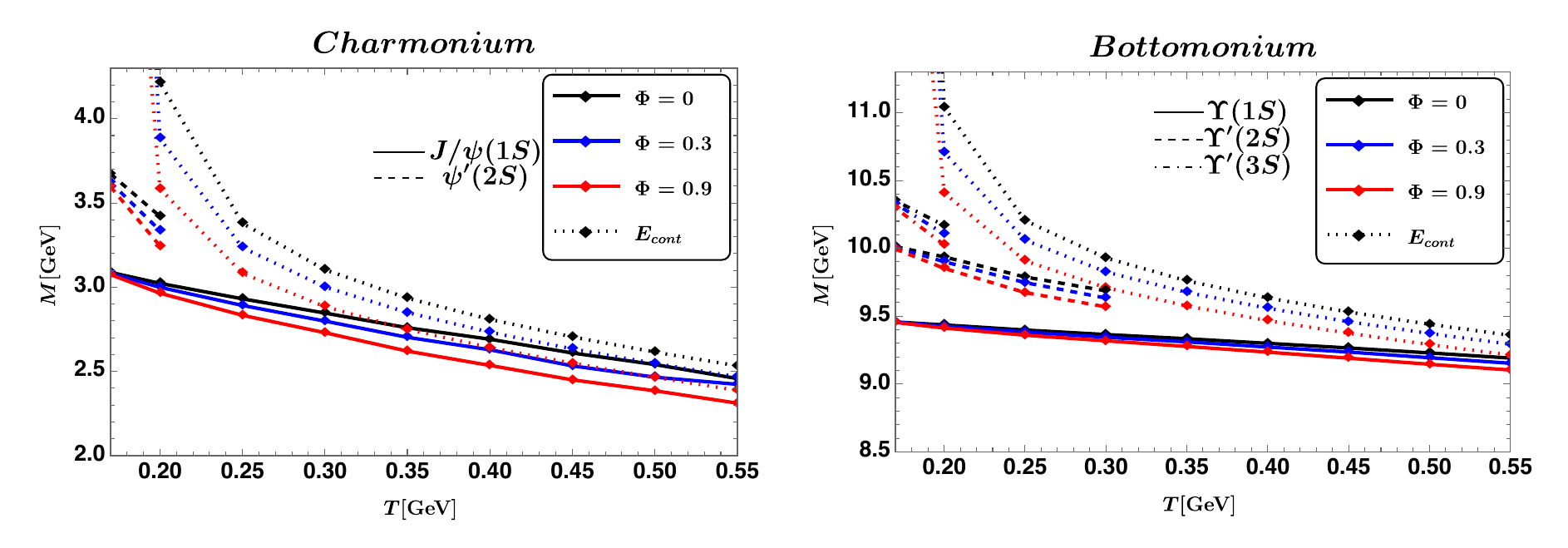}     
	\caption{The in-medium masses of ground and excited states of
		charmonium (left) and bottomonium (right) for different values of $ \Phi $. Dotted lines show the
		continuum threshold energy, 
		($E_{\rm cont}= {\Re V}(r\rightarrow \infty ) + 2m_{Q} $.)
	}
	\label{MEcont}
	\vspace{-0.7cm}
\end{figure}
%

Using the obtained potential, we compute the spectral functions of heavy quarkonia 
to assess the impact of bulk viscosity on the properties of quarkonia 
via the Fourier space method developed in ref.~\cite{Burnier:2007qm}. 
We first compute the unequal-time point-split meson-meson correlator, $C^{>} (t;{\bf x},{\bf x'})$, which satisfy the following Schr\"odinger equation,
\begin{equation}
	\,\,\,	\left[2m_Q-\frac{\nabla^{2}_{r}}{m_Q}+\frac{l(l+1)}{m_Qr^2}
	+{\Re }V(r,T,\Phi)\mp i | {\Im}V(r,T,\Phi)|\right]C^{>} (t;{\bf x},{\bf x'})=i\partial_t C^{>} (t;{\bf x},{\bf x'}), 
	\label{schreq}
\end{equation}
in which we use the potential given by
eqs.~(\ref{eq:pot-realphi})  and~(\ref{eq:pot-imaginary}).
The vector-channel spectral function is obtained through Fourier transformation of
$C^{>} (t;{\bf x},{\bf x'})$ as 
\begin{equation}
	\rho^V({\omega})=\lim_{{\bf x},{\bf x'}\rightarrow 0}\frac{1}{2}\tilde{C}(\omega;{\bf x},{\bf x'}). 
\end{equation}
To extract physical properties from the obtained spectral functions, 
we fit each peak of the in-medium spectral functions with a skewed Breit-Wigner form~\cite{Lafferty:2019jpr}.

\section{ Results }

\begin{figure}
	\includegraphics[width=14cm]{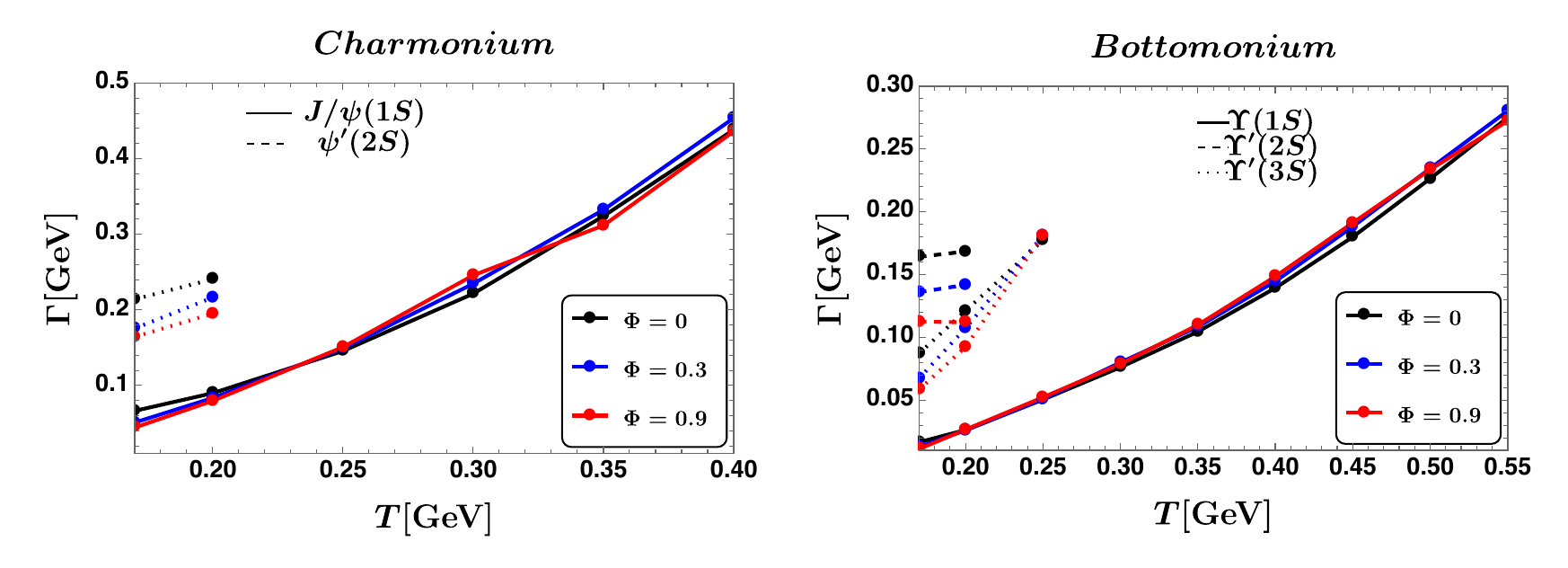}     
	\caption{Thermal widths of the ground and excited states of  charmonium (left) and bottomonium (right) as a function of temperature for different values of $ \Phi $.
	}
	\label{DecayW}
	\vspace{-0.6cm}
\end{figure}

Let us now discuss the results. In figure~\ref{SFvsw},  we plot the spectral functions of 
the charmonium and bottomonium 
for different values of $\Phi$. 
We observe that the peaks of the spectral functions shift towards lower values of $ \omega $ as $ \Phi $ increases.
Figure~\ref{MEcont} shows that the in-medium masses, $M$, of the each state and 
threshold energy, $E_{\rm cont}$,  
decreases as a function of $T$ and $\Phi$. 
Looking at figure~\ref{DecayW}, we find that the decay widths of the ground states of charmonium and bottomonium do not show monotonic tendency 
as a function of $\Phi$. However, the decay widths of the excited states are clearly decreasing functions of $\Phi$.
Namely, it is important to recognize that ground  and excited states exhibit distinct responses to the influence of bulk viscosity, with excited states displaying a heightened sensitivity to this non-equilibrium effect.
From an experimental perspective, it becomes intriguing to investigate the collision energy dependence of the nuclear modification factor $R_{AA}$ for both excited and ground states. In particular, should there be an enhancement in the bulk viscosity near the critical point, it could potentially give rise to a non-monotonic behavior in the $R_{AA}$ of excited states.
%

We also analyze the effect of bulk viscous correction on the
$\psi^{\prime}$ to $J/\psi$ ratio, 
following the method described in ref.~\cite{Lafferty:2019jpr}. 
The ratio is obtained from the dilepton production rate, $R_{l\bar{l}}$, which is computed using spectral functions and is based on a projection process implemented by replacing the in-medium peaks with the delta-function peaks~ \cite{Lafferty:2019jpr} 
\begin{equation}
	\frac{N_{\psi^{\prime}}}{N_{J/\psi}}
	=
	\frac{	R_{l\bar{l}}^{\psi^{\prime}}}{R_{l\bar{l}}^{J/\psi}}\cdot \frac{M_{\psi^{\prime}}^2|\psi_{J/\psi}(0)|^2}{M_{J/\psi}^2|\psi_{\psi^{\prime}}(0)|^2}.
	\label{eq:ratio}
\end{equation}
We use the freeze-out temperature to relate the ratio with collision energies, $\sqrt{s_{NN}} $, which are fitted to reproduce the particle yields~\cite{Andronic:2017pug} and compute the $\psi^{\prime}/ J/\psi$ ratio over a range of $\sqrt{s_{NN}}$. 
\begin{figure}
	\begin{center}
		\includegraphics[width=10cm]{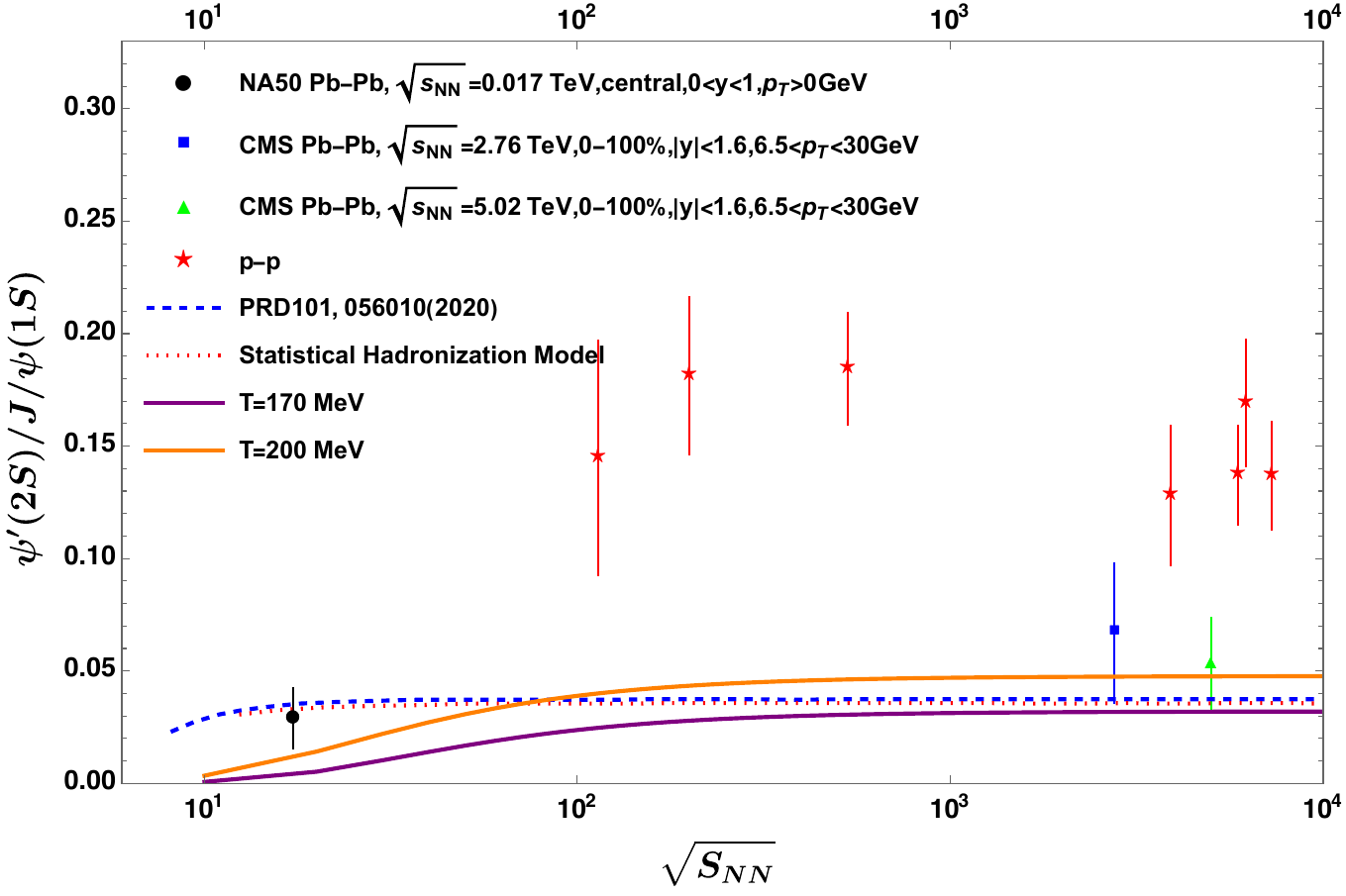}
		\caption{
			$ \psi^{\prime}/ J/\psi$ ratio as a function of $ \sqrt{s_{NN}} $. The red ($T=170$) and purple ($T=200$ MeV) solid lines present this work. 
			The blue dashed and red dotted lines show 
			Gauss law potential model~\cite{Lafferty:2019jpr} 
			and statistical hadronization model~\cite{Andronic:2017pug}. Different data points are taken from ref.~\cite{Lafferty:2019jpr}.  
		}
		\label{ratio}
	\end{center}
	\vspace{-0.8cm}
\end{figure} 
In Figure~\ref{ratio}, we plot the computed ratio at the freezing temperatures, $170$ MeV and $200$ MeV.  Our results are in reasonable agreement with the statistical hadronization model, Gauss law potential model, and  
experimental data  from the CMS Collaboration.
We also examine the effect of bulk viscous correction $\Phi$ on the 
$ \psi^{\prime}/ J/\psi$. See ref.~\cite{Thakur:2021vbo} for details.
\section{Conclusion}
We investigated the influence of bulk viscosity on quarkonium properties and experimental observable in heavy-ion collisions. We extracted the in-medium masses and decay widths of quarkonium states from the spectral functions. 
The in-medium masses of quarkonium states decrease with increasing bulk viscous correction. 
While the decay widths of the ground states do not show any particular trend, 
the decay widths of the excited states decrease with increasing bulk viscous correction.
This feature can be helpful in identifying the QCD critical point where the bulk viscous effect is expected to increase. It would be interesting to compare the collision energy dependence of the nuclear modification factor, $R_{AA}$, of the ground and excited states.
\acknowledgments
LT would like to thank the organizers of the  HADRON 2023 Conference for the opportunity to give a parallel session talk.
LT and YH were supported by National Research Foundation (NRF) funded by the Ministry of Science of Korea with Grant No.~2021R1F1A1061387(LT) and 2020R1F1A1076267(YH).

\end{document}